\documentclass[prb, twocolumn, showpacs]{revtex4}         

\usepackage{amsmath, amssymb, graphicx, subfigure, bm}

\DeclareMathOperator{\const}{const}
\DeclareMathOperator{\re}{Re}
\DeclareMathOperator{\im}{Im}

\newcommand{\figwidth}{0.48\textwidth}

\begin{document}

\title{Macroscopic Quantum Tunneling in Small Antiferromagnetic Particles:
Effects of a Strong Magnetic Field}

\author{B.~A.~Ivanov}
\email{bivanov@i.com.ua}

\author{V.~E.~Kireev}
\email{kireev@imag.kiev.ua}

\affiliation{Institute of Magnetism NAS of Ukraine, 36-B Vernadskii avenue,
03142 Kiev, Ukraine}

\date\today

\begin{abstract}
  We consider an effect of a strong magnetic field on the ground state and
  macroscopic coherent tunneling in small antiferromagnetic particles with
  uniaxial and biaxial single-ion anisotropy.  We find several tunneling regimes
  that depend on the direction of the magnetic field with respect to the
  anisotropy axes.  For the case of a purely uniaxial symmetry and the field
  directed along the easy axis, an exact instanton solution with two different
  scales in imaginary time is constructed.  For a rhombic anisotropy the effect
  of the field strongly depends on its orientation: with the field increasing,
  the tunneling rate increases or decreases for the field parallel to the easy
  or medium axis, respectively.  The analytic results are complemented by
  numerical simulations.
\end{abstract}


\maketitle

During the last decade macroscopic coherent quantum tunneling between two
equivalent classical ground states in macroscopic or, to be more precise,
mesoscopic magnetic systems became an object of intense experimental and
theoretical investigations, see for a review Refs.\cite{ChudnTej98b, QTM95}. In
the physics of magnetism such systems are, for instance, small magnetic
particles, magnetic clusters, and high-spin molecules.  The interest in
tunneling effects is associated with the possibility of using these systems as
potential elements for quantum computers.  Initially, the calculations of
tunneling effects were carried out for ferromagnets.\cite{Chudn79,
  ChudnGunter88prb} However it happened that the tunneling effects were
experimentally observed by the resonant absorption of electromagnetic waves for
antiferromagnetic ferritin particles.\cite{Aw+92} The effect of the magnetic
field on the tunneling probability in ferritin particles have been
experimentally studied, see Ref.\cite{Tej+97} According to theoretical
estimations,\cite{BarbChudn90, KriveZasl90} the level spitting in
antiferromagnets is stronger than in ferromagnets and the effects can be
observed at higher temperatures.  Antiferromagnets are more convenient for
experimental investigations of tunneling effects.

The interest in tunneling in compensated antiferromagnets has been renewed after
the synthesis of high-spin molecules with antiferromagnetic coupling of spins,
so-called ferric wheels, such as Fe$_6$, Fe$_{10}$, Cr$_8$, see, for instance,
Refs.\cite{Gatt+94, CornJansAffr99, Waldmann+02} The interference effects caused
by the gyroscopic terms in the $\sigma$-model approximation were of main
interest.\cite{GolPopkov95} For compensated antiferromagnets such effects can
arise due to an external magnetic field\cite{GolPopkov95, ChiolLoss98} or a
certain kind of the Dzyaloshinskii-Moriya interaction.\cite{IvKir02} Chiolero
and Loss\cite{ChiolLoss98} have also found a type of interference effect for
such antiferromagnets placed in a strong magnetic field directed perpendicular
to the easy axis of the antiferromagnetic particle.  This effect caused by the
imaginary part of the fluctuation determinant together with the direct
contribution of the field to the imaginary part of the Euclidean action produces
the periodic dependence of the level splitting on the field strength.  Then Hu
et al.\cite{Hu+00} have investigated these effects for a more general case of
antiferromagnets with rhombic anisotropy.  They have considered the case of the
field directed along the hard axis and found the behavior common to that in
Ref.\cite{ChiolLoss98}. They also suggested that such properties are universal
for a wide class of antiferromagnets subject to a strong magnetic field.

In this paper we consider the effects of coherent quantum tunneling for a
compensated antiferromagnetic particle with rhombic anisotropy placed in the
magnetic field.  In the section~\ref{s:model} we describe the model used for the
evaluation of tunnel splitting and analyze the static energy of the particle for
three different orientations of the magnetic field along the crystalline axes.
The section~\ref{s:instanton} is devoted to the calculation of the level
splitting of the ground states in the instanton approximation for the
orientations of the field along the easy, middle and hard axis separately and
for relevant ranges of the field strength.  The section~\ref{s:numerics}
contains a qualitative comparison of analytical results with data of numerical
calculations for level splitting and a direct comparison for the Euclidean
action.  The final section~\ref{s:conclusion} resumes obtained results and gives
a short survey of experiments where the predicted effects could be observed.  We
use the standard semiclassical analysis based on the instanton technique applied
to the nonlinear $\sigma$-model, as well as the direct numerical diagonalization
of the corresponding quantum spin Hamiltonian.  We demonstrate that the behavior
of the level splitting is highly sensitive to the orientation of the magnetic
field.  In fact, three different scenarios for the orientation of the field
along the symmetry axes are found.

\section{Nonlinear $\sigma$-model for antiferromagnets in a magnetic field 
\label{s:model}}

We start from the Hamiltonian describing a magnetic particle with an even number
$N$ of magnetic ions with the spin $S$ and coupled with the nearest-neighbor
antiferromagnetic exchange interaction $J$.  We also assume that a single-ion
anisotropy with rhombic symmetry and the magnetic field $\bm{H}$ are present.
The macroscopic Hamiltonian of the system can be written as
\begin{multline} \label{Hamiltonian}
\mathcal{H} =
J \sum_{<\alpha\beta>} \bm{S}_\alpha \bm{S}_\beta +
B_u \sum_\alpha \left[ \left(S^x_\alpha\right)^2 + 
\left(S^y_\alpha\right)^2 \right] \\
+ B_p \sum_\alpha \left(S^y_\alpha\right)^2 -
g \mu_B \sum_\alpha \bm{H} \cdot \bm{S}_\alpha \;.
\end{multline}
Here, $\bm{S}_\alpha$ is the spin at the $\alpha$th site, $g \approx 2$ is the
Land\'e factor, and $\mu_B$ is the Bohr magneton.  The first term describes the
isotropic exchange interaction $J > 0$, and the summation in this term is
extended over the pairs of nearest neighbors.  The second and third terms give
the simplest form of a rhombic single-ion anisotropy; $B_u$ and $B_p$ are the
constants of the uniaxial anisotropy and the anisotropy in the basal plane $xy$,
respectively.  We assume that $B_u > 0$ and $B_p > 0$.  Thus, $z$ is the easy
axis and $x$, $y$ are the medium and hard axes, respectively.  The last term
corresponds to the coupling spins with the magnetic field.

We will analyze the system under the assumption that all spin pairs are
equivalent and have the same coordination number $z$.  It is true for spin
dimers ($z = 1$), spin wheels ($z = 2$), and can be a good approximation for
mesoscopic three-dimensional crystalline particles with $N \gg 1$, in which the
surface variation of parameters can be neglected.  Assume that the Zeeman energy
$g \mu_B H S$, the anisotropy energy $B_p S^2$, and $B_u S^2$ are much smaller
than the exchange energy $J z S^2$.  The classical approximation gives that the
spins in each sublattice are parallel.  This assumption is a necessary condition
for using the $\sigma$-model as a model for describing both classic and quantum
dynamics of antiferromagnets.\cite{ChiolLoss98} For the semiclassical dynamics,
especially for tunneling, it is convenient to use the Euclidean formulation,
which is based on the introduction of imaginary time $\tau = it$.

The Euclidean action for the $\sigma$-model that corresponds to the microscopic
Hamiltonian~\eqref{Hamiltonian} can be written as
\begin{multline} \label{action}
\mathcal{A}_E[\bm{l}(\tau)] =
\int^{+\infty}_{-\infty} \mathcal{L}_E [\theta(\tau),\
\phi(\tau) ] \; d\tau  = \\
N \int^{+\infty}_{-\infty} d\tau \left\{\frac{\hbar^2}{4J z} \left[
\dot{\bm{l}}^2 -
2i \gamma\bm{H} \cdot (\bm{l} \times \dot{\bm{l}}) \right] +
w_a(\bm{l}) \right\} \;,
\end{multline}
where $\gamma = g \mu_B / \hbar$ is the gyromagnetic ratio, $N$ is the number of
magnetic ions in both sublattices and the overdot denotes the derivative with
respect to the imaginary time $\tau$.

In this approach the total spin $\bm{S}_{tot}$ of the particle becomes a slave
variable, and it is determined through the unit N\'eel vector $\bm{l}$ and its
derivative $\dot{\bm{l}}$ with respect to $\tau$:
\begin{equation} \label{total_spin}
\bm{S}_{tot} =
\frac{\hbar N}{J z} \left\{\gamma \left[ \bm{H} - \bm{l}(\bm{H} \cdot \bm{l}) \right]
+ i\left(\bm{l} \times \dot{\bm{l}} \right) \right\} \;.
\end{equation}

The function $w_a(\bm{l})$ is an effective energy of anisotropy per spin with
renormalization caused by the external field
\begin{equation} \label{anisotropy}
w_a(\bm{l}) =
B_u S^2 (l_x^2 + l_y^2) + B_p S^2 l_y^2 +
\frac{(g \mu_B)^2}{4J z}(\bm{H} \cdot \bm{l})^2 \;.
\end{equation}

In accordance with the general rules of the semiclassical approximation
formulated in the instanton language the amplitude of the tunnel transition from
the state $\bm{l} = \bm{l}^-$ to $\bm{l} = \bm{l}^+$ is proportional to
$\exp(-\mathcal{A}_E / \hbar)$, where the value of $\mathcal{A}_E$ is calculated
from appropriate equations of motion under the boundary conditions:
$\bm{l}(\tau) \to \bm{l}^-$ at $\tau \to -\infty$ and $\bm{l}(\tau) \to
\bm{l}^+$ at $\tau \to +\infty$.  Tunnel splitting in the so-called dilute
instanton gas approximation is determined by the modulus of the sum of such
amplitudes calculated along all instanton trajectories with a minimal value of
$\re\mathcal{A}_E$.

It is convenient to introduce a polar parametrization for the unit vector
$\bm{l}$
\begin{equation} \label{parameterization}
l_z = \cos\theta, \quad
l_x = \sin\theta \cos\phi, \quad
l_y = \sin\theta \sin\phi \;.
\end{equation}
The Lagrangian from Eq.~\eqref{action} in such a parametrization takes the form
\begin{multline} \label{Lagrangian}
\mathcal{L}_E [\theta(\tau), \phi(\tau)] = 
N w_a(\theta, \phi) \\
+ \frac{\hbar^2 N}{4J z} \biggl(
\dot{\theta}^2 + \dot{\phi}^2 \sin^2\theta - 2i \gamma
\biggl\{ \dot{\theta} (H_y \cos\phi - H_x \sin\phi) \\ +
\dot{\phi} \left[ H_z \sin^2\theta -
(H_x \cos\phi + H_y \sin\phi) \sin\theta \cos\theta \right] \biggr\} \biggr) \;.
\end{multline}
Here we omitted the term with a full derivative proportional to $N S \dot{\phi}$
because it does not create the interference effects for the tunneling between
opposite points of the unit sphere for a fully compensated antiferromagnet.

At zero magnetic field instanton solutions that give minima of the
action~\eqref{action} with the boundary conditions $\bm{l}(\pm\infty) \to
\pm\hat{\bm{e}}_z$ can be easily found.  The above-mentioned conditions take the
form $\theta(-\infty) = 0$, $\theta(+\infty) = \pi$ with arbitrary values for
$\phi(\pm\infty)$.  The instanton solutions correspond to the rotation of
$\bm{l}$ in the symmetry planes of the system, i.e., $\phi = \pi k / 2$, where
$k$ is integer, and
\begin{equation} \label{instanton}
\cos\theta = \tanh (\overline{\omega} \tau), \quad
\sin\theta = \frac{\sigma}{\cosh (\overline{\omega} \tau)} \;,
\end{equation}
where $\sigma = \pm 1$ is the polarization of the instanton.  The ``frequency''
$\overline{\omega}$ is $2S \sqrt{J z B_u} / \hbar$ for the instanton paths going
through the medium axis $\phi = 0, \pi$ and $\overline{\omega} = 2S \sqrt{J z
  (B_u + B_p)} / \hbar$ for the paths going through the hard axis $\phi = \pm
\pi / 2$.  The Euclidean action calculated on these instantons is simply
$\mathcal{A}_E = \hbar N (\hbar \overline{\omega} / J z)$.  Following the
saddle-point approximation that corresponds to the dilute instanton gas we take
into account only paths with a minimal real part of the Euclidean action, i.e.
with the smallest $\overline{\omega} = 2S \sqrt{J z B_u} / \hbar$.  Thus, at
zero magnetic field the only instanton pair with $\sigma = \pm 1$ and the
rotation of $\bm{l}$ in the plane $xz$ contributes to the tunneling.  The
presence of the magnetic field drastically changes the features of the spin
tunneling between two different classical ground states in the antiferromagnetic
particle.

First of all we consider the influence of the magnetic field on the static
classical properties of the antiferromagnetic particle.  We will use the
expression~\eqref{anisotropy} and restrict ourselves to the cases when the field
is parallel to one of the symmetry axes $x$, $y$, or $z$.  In these cases the
field simply renormalizes the value of the appropriate anisotropy constants.

The influence of the field on the ground state is essential when the field is
directed along the easy axis $z$.  In this case the renormalized constant of the
uniaxial anisotropy $B_u(H)$ can change its sign, $B_u(H) = B_u (1 - H^2 /
H_u^2)$, where
\begin{equation} \label{SF-field}
H_u =
\frac{2 S \sqrt{J z B_u}}{g \mu_B}
\end{equation}
is the field of the familiar spin-flop phase transition in the classical theory
of antiferromagnetism.  When $H > H_u$ the states with $\bm{l} \parallel \pm
\hat{\bm{e}}_z$ become unstable and the ground states have $\bm{l}$ parallel to
$\pm \hat{\bm{e}}_x$.

When $\bm{H}$ is parallel to the medium axis $x$, the constant $B_u$ increases,
$B_u(H) = B_u + (g \mu_B H)^2 / (4 J z S^2)$ and $B_p$ decreases as $B_p(H) =
B_p - (g \mu_B H)^2 / (4 J z S^2)$, with the growth of the field.  Thus, the
field does not effect on the ground state, but a strong enough field can change
the type of the axes $x$ and $y$ in the basal plane.  When $H > H_p$, where
\begin{equation} \label{c-field}
H_p =
\frac{2 S \sqrt{J z B_p}}{g \mu_B} \;,
\end{equation}
the axis $y$ becomes an easy direction in the basal plane (medium axis) and the
axis $x$ is a hard one.

Finally, if the magnetic field is directed along the hard axis $y$, both
anisotropy constants increase as $B_{u, p}(H) = B_{u, p} + (g \mu_B H)^2 / (4 J
z S^2)$.  In this case, the field does not change the ground state, as well as
the type of the axes.

The naive substitution of the renormalized constants into the
expressions~\eqref{instanton} for an instanton and the Euclidean action
calculated on it leads to a wrong prediction that when the field is directed
along the easy axis $\mathcal{A}_E \to 0$ at $H \to H_u$, and for the field
along the medium axis the values of $\mathcal{A}_E$ on the two classes of
trajectories [$\phi = \pi k$ and $\phi = (2k + 1) \pi / 2$] become equal at $H =
H_p$.  As we will show below, both suggestions are wrong.

In addition to the abovementioned renormalizations of the anisotropy constants,
the field changes the dynamics of the vector $\bm{l}$ and leads to the
appearance of a gyroscopic term linear in $d\bm{l} / d\tau$ in the Lagrangian.
The role of gyroscopic terms for tunneling in antiferromagnets has been
discussed in Refs.\cite{GolPopkov95, ChiolLoss98}.  The authors of these papers
have shown that the gyroscopic term caused by the magnetic field can create the
imaginary part of the Euclidean action $\mathcal{A}_E$ that is a linear function
of the magnetic field.  As we have shown, similar effects can also arise due to
the Dzyaloshinskii-Moriya interaction.\cite{IvKir02} The imaginary part of the
Euclidean action $\mathcal{A}_E$ can lead to the interference effects and to the
oscillations of the tunnel splitting as a function of the magnetic field.  We
will show below that the gyroscopic term produces a dynamic renormalization of
the real part of $\mathcal{A}_E$, which is quadratic in the magnetic field.  It
can completely suppress the static contribution to $\mathcal{A}_E$ coming from
the renormalization of the anisotropy constants.


\section{Instanton solutions \label{s:instanton}}

In order to describe macroscopic quantum tunneling between two classical states
$\bm{l} = \hat{\bm{e}}_z$ and $\bm{l} = -\hat{\bm{e}}_z$ in the saddle-point
approximation, it is necessary to find instanton solutions of the two
Euler-Lagrange equations for the Euclidean action~\eqref{action} for the
independent variables $\theta(\tau)$ and $\phi(\tau)$.  Using the identities
$(\delta / \delta\theta) \int d\tau \bm{H} \cdot (\bm{l} \times \dot{\bm{l}}) =
-2 \dot{\phi} (\bm{H l}) \sin\theta$ and $(\delta / \delta\phi) \int d\tau
\bm{H} \cdot (\bm{l} \times \dot{\bm{l}}) = 2 \dot{\theta} (\bm{H l})
\sin\theta$ that can be obtained through the variation of the
Lagrangian~\eqref{Lagrangian}, it is convenient to write down the Euler-Lagrange
equations with the magnetic field $\bm{H}$ directed along an arbitrary symmetry
axis
\begin{widetext}
\begin{subequations} \label{eul}
\begin{align}
\ddot{\theta} -
\left\{\omega_u^2 + \dot{\phi}^2 + \gamma^2 (H_x^2 - H_z^2) +
\left[\omega_p^2 + \gamma^2 (H_y^2 - H_x^2) \right] \sin^2\phi \right\}
\sin\theta \cos\theta & =
-2i \dot{\phi} \gamma (\bm{H l}) \sin\theta \;, \label{eul-theta} \\
\ddot{\phi} \sin^2\theta + 2\dot{\phi} \dot{\theta} \sin\theta \cos\theta -
\left[\omega_p^2 + \gamma^2 (H_y^2 - H_x^2) \right]
\sin^2\theta \sin\phi \cos\phi & =
\phantom{-} 2i \dot{\theta} \gamma (\bm{H l}) \sin\theta \;. \label{eul-phi}
\end{align}
\end{subequations}
\end{widetext}
Here $\omega_u = \gamma H_u = 2 S \sqrt{2J z B_u} / \hbar$ and $\omega_p =
\gamma H_p = 2 S \sqrt{2J z B_p} / \hbar$.  The terms in the right-hand side are
responsible for the gyroscopic dynamics of the vector $\bm{l}$.  Note that we
consider the field directed along one of the crystalline axes only, so only one
of the components $\bm{H}$ in the system~\eqref{eul} is nonzero.

It is important to note that in the case $\bm{H} \neq 0$ simple planar solutions
such as $\phi = \pi k / 2$ may not exist.  If such solutions are absent, the
full system~\eqref{eul} is equivalent to the Lagrange equations for a mechanical
system with two degrees of freedom.  To integrate such a system, the existence
of two independent integrals of motion is necessary.  For the general case
$\omega_u \neq 0$ and $\omega_p \neq 0$ only one first integral is known
\begin{equation} \label{e_integral}
\mathcal{E} =
\dot{\theta}^2 + \dot{\phi}^2 \sin^2\theta -
\omega_u^2 \sin^2\theta - \omega_p^2 \sin^2\theta \sin^2\phi \;,
\end{equation}
with the value $\mathcal{E} = 0$ for separatrix solutions we are interested in.
For this reason the system~\eqref{eul} cannot be solved analytically.  However,
approximate solutions can be constructed for all cases of interest; see a
detailed consideration in the following sections.

\subsection{Field parallel to the easy axis}

We will start from the case of the field parallel to the easy axis, for which
the gyroscopic terms in the right-hand side of Eqs.~\eqref{eul} are independent
of $\phi$.  The field does not violate the rotational symmetry around the easy
axis, and the model with an isotropic basal plane ($\omega_p = 0$) has a
physical meaning for this case.  Its analysis leads to instructive results and
we give it completely.

If $\omega_p = 0$, the system~\eqref{eul} has one more integral of motion, which
can be written as
\begin{equation} \label{m_integral}
\Omega =
(\dot{\phi} - i \gamma H) \sin^2\theta =
\const \;.
\end{equation}
Using the two integrals of motion \eqref{e_integral} and \eqref{m_integral}, we
can simplify the system to the ordinary differential equation for $\theta(\tau)$
only
\begin{equation} \label{em_integral}
\dot{\theta}^2 - \omega_u^2 \sin^2\theta -
\frac{\Omega^2}{\sin^2\theta} = \mathcal{E}
\end{equation}
and integrate it exactly.  The instanton solutions with appropriate boundary
conditions $\theta \to 0, \pi$ and $\dot{\theta} \to 0$ at $\tau \to \pm\infty$
correspond to the values of the integrals $\Omega = 0$, $\mathcal{E} = 0$.
Thus, the solution is $\phi = i \gamma H (\tau - \tau_1)$, where $\tau_1$ is an
arbitrary constant.  Eq.~\eqref{em_integral} simplifies to $\dot{\theta}^2 =
\omega_u^2 \sin^2\theta$, and its solution is described by the
formula~\eqref{instanton} with $\overline{\omega} = \omega_u$.  In terms of the
vector $\bm{l}$, the instanton solution for $\bm{H}$ parallel to the easy axis
and $\omega_p = 0$ takes the form
\begin{subequations} \label{rotator-instanton}
\begin{align}
l_x & = \sigma \frac{\cosh[\gamma H(\tau - \tau_1)]}
{\cosh[\overline{\omega}_0(\tau - \tau_0)]} \;, \\
l_y & = i\sigma \frac{\sinh[\gamma H(\tau - \tau_1)]}
{\cosh[\overline{\omega}_0(\tau - \tau_0)]} \;, \\
l_z & =
\tanh[\overline{\omega}_0(\tau - \tau_0)] \;.
\end{align}
\end{subequations}
This solution has correct instanton asymptotics $l_{x, y} \to 0$ and $l_z \to
\pm 1$ at $\tau \to \pm \infty$ for all $H$ in the range $\gamma H < \omega_u$,
which corresponds to $H < H_u$.  It is valid in the full region of stability
of the states $\bm{l} = \pm \hat{\bm{e}}_z$.  Though the constant of effective
anisotropy $B_u(H)$ changes from $B_u$ at $H = 0$ to $0$ at $H = H_u$, the
Euclidean action calculated on the solution~\eqref{rotator-instanton} does not
depend on $H$, its real part is the same as at $H = 0$ and the imaginary part is
zero.  Thus, in the uniaxial case $B_p = 0$ and $H < H_u$ the action is
\begin{equation} \label{rotator-action}
\mathcal{A}_E = 2 \hbar N S \sqrt{\frac{B_u}{J z}} \;.
\end{equation}

This simple example shows that the gyroscopic term can essentially suppress the
static renormalization of the anisotropy constant.  For the case $\omega_p = 0$
the renormalization is completely compensated by the gyroscopic term and
$\mathcal{A}_E$ is real and does not depend on the field.

For the case $\omega_p = 0$ an exact solution of the problem of small
fluctuations around the classical instanton solution is possible to construct,
and the preexponential factor can be explicitly written.  To do so, we introduce
small perturbations $\vartheta$ and $\mu$ as
\begin{subequations}\label{deviations}
\begin{align}
\theta(\tau) & = \theta_0(\tau) + \vartheta \;, \\
\phi(\tau)   & = \phi_0(\tau) + \frac{\mu}{\sin\theta_0(\tau)} \;,
\end{align}
\end{subequations}
where $\theta_0(\tau)$ and $\phi_0(\tau)$ correspond to the instanton solution
given by Eqs.~\eqref{rotator-instanton}.  The variational part of the action is
$\mathcal{A}_E - \mathcal{A}_E^{(0)}$, where $\mathcal{A}_E^{(0)}$ is calculated
on the unperturbed instanton solution.  It can be written as a sum of two
independent terms $\mathcal{A}_E - \mathcal{A}_E^{(0)} = \delta^2
\mathcal{A}_E^{(\theta)} + \delta^2 \mathcal{A}_E^{(\phi)}$ with the decoupled
degrees of freedom $\vartheta$ and $\mu$.  Both terms have the same form
\begin{equation}\label{action_variation}
\delta^2 \mathcal{A}_E^{(\alpha)} =
\frac{N \hbar^2 \overline{\omega}}{4J z}
\int_{-\infty}^{+\infty}d\xi \, (f^{(\alpha)}, \widehat{M} f^{(\alpha)}) \;,
\end{equation}
where $\xi = \overline{\omega} \tau$, $f^{(\alpha)}$ is $\vartheta$ or $\mu$ for
$\delta^2 \mathcal{A}_E^{(\theta)}$ or $\delta^2 \mathcal{A}_E^{(\phi)}$,
respectively, and $\widehat{M}$ is the linear operator
\begin{equation} \label{BOSH}
\widehat{M} =
-\frac{d^2}{d\xi^2} + 1 - \frac{2}{\cosh^2\xi} \;.
\end{equation}
This operator frequently appears in scattering problems associated with soliton
theory, so its properties are well studied.  The full set of its eigenvalues and
normalized eigenfunctions is
\begin{subequations}\label{BOSH-spectrum}
\begin{align}
\widehat{M} f_0 & = 0, \quad &
f_0 & = \frac{1}{\sqrt{2} \cosh\xi} \;, \label{BOSH-spectrum-loc} \\
\widehat{M} f_k & = (1 + k^2) f_k, \quad &
f_k & = \frac{(\tanh\xi - i k)e^{-i k \xi}}{\sqrt{L (1 + k^2)}}
\label{BOSH-spectrum-cont} \;.
\end{align}
\end{subequations}
The mode $f_0$ is a localized eigenfunction and the modes with $f = f_k$ form a
continuous spectrum.

It is worthwhile to note that in contrast to other problems of macroscopic
quantum tunneling, for the case of antiferromagnets with $\omega_p = 0$ the zero
mode $f_0$ is present for \emph{both} kinds of fluctuations.  The second zero
mode is caused by the exact rotational symmetry around the axis $z$.  The full
preexponential factor for this problem is a \emph{square} of the usual
fluctuation determinant $D = \sqrt{\mathcal{A}_E^{(0)} / (2 \pi \hbar)}$, and
the tunnel splitting of the lowest levels $\Delta$ takes the form
\begin{equation}\label{Delta}
\Delta =
C \hbar \omega_u \left(\frac{\mathcal{A}_E^{(0)}}{2 \pi \hbar}
\right) \exp\left(-\mathcal{A}_E^{(0)} / \hbar \right) \;,
\end{equation}
where $C$ is a numerical constant of order of unity.  The additional large
factor $\sqrt{\mathcal{A}_E^{(0)} / (2 \pi \hbar)}$ is a consequence that the
instanton solution \eqref{rotator-instanton} in the case $\omega_p = 0$ contains
two (not one, as usually) continuous parameters $\tau_1$ and $\tau_0$.  Note
that the level splitting $\Delta$ does not depend on the magnetic field even if
the fluctuation determinant is taken into account.  This result can be explained
using exact quantum-mechanical arguments.  For uniaxial system the Hamiltonian
of the system commutes with the $z$projection of the total spin $\hat{S}_z$, and
the eigenstates of the problem can be characterized by definite values $S_z = 0,
\pm1, \pm2, \ldots$ The two lowest levels form a doublet with $S_z = 0$ and the
magnetic field does not influence on them.  For this model the spin-flop
transition at the field $H = H_u$ corresponds to the change of the value of
$S_z$ for the lowest level from $S_z = 0$ to the value $S_z = 1$ or higher.
Then, the growth of the magnetic field leads to the growth of $S_z$ and a
sawlike dependence $\Delta(H)$ appears.  Since these effects are not associated
with tunneling, we will not consider them in the following, and restrict
ourselves only to the region $H < H_u$.

Now we consider a more general case $\omega_p \neq 0$.  It is clear that at
finite nonzero $\omega_p$ the additional factor $(\mathcal{A}_E / 2 \pi
\hbar)^{1/2}$ is absent, and the preexponential factor is smaller than that for
the uniaxial case $\omega_p = 0$.  The levels cannot be characterized by the
quantum number $S_z = 0, \pm1, \pm2, \ldots$, and their splitting becomes
dependent on the magnetic field.

The instanton solutions with $\theta = \theta(\tau)$, $\phi = 0$ exist at $H =
0$.  It can be expected that at $\gamma H \ll \omega_u$ the value of $\phi$ or,
more accurately, the appropriate projection of the vector $\bm{l}$, $l_y \simeq
\sin\phi \sin\theta_0$ is small, and the function $\theta(\tau)$ can be given by
the solution similar to Eq.~\eqref{rotator-instanton}.  If $\omega_p \gg
\omega_u$, the out-of-plane components of $\bm{l}$ are small, i.e. $|l_y| \ll
1$.  Assuming that $|l_y| \ll 1$ for all values of parameters, the variational
technique can be applied for evaluating the action.  Choosing as a trial
function $\cos\theta = \tanh [\overline{\omega} (\tau - \tau_1)]$ with some
parameter $\overline{\omega}$ that have to be found later, we rewrite
Eq.~\eqref{eul-phi} in the linear approximation in $l_y$ as
\begin{equation} \label{ea-operator}
\left(\widehat{M} + \epsilon \right) l_y(\xi) =
\frac{2i \gamma H}{\overline{\omega}} \;
\frac{\sinh\xi}{\cosh^2\xi} \;.
\end{equation}
Here $\epsilon = (\omega_p / \overline{\omega})^2$, $\xi = \overline{\omega}
\tau$, and $\widehat{M}$ is the linear operator~\eqref{BOSH}, introduced above,
with non-negative eigenvalues.  Since $\epsilon > 0$, the operator $\widehat{M}
+ \epsilon$ is positively defined.  It has an inverse operator which can be
written through the ordinary bra and ket notation of eigenfunctions as
\begin{equation} \label{BOSH-inverse}
\frac{1}{\widehat{M} + \epsilon} =
\frac{|f_0\rangle \langle f_0|}{\epsilon} +
\sum_k \frac{|f_k\rangle \langle f_k|}{1 + \epsilon + k^2} \;,
\end{equation}
where the summation is extended over the continuous spectrum.  Since $\langle
\sinh \xi / \cosh^2 \xi, \; f_0 \rangle = 0$, the formal solution of
Eq.~\eqref{ea-operator} does not contain a term with $1 / \epsilon$.  The
solution $l_y(\xi)$ is determined by the summation over the states of the
continuous spectrum $f_k$ only.  After simple calculations the Euclidean action
as a function of the trial parameter $\overline{\omega}$ takes the form
\begin{multline} \label{ea-action0}
\mathcal{A}_E(\overline{\omega}) =
\frac{\hbar^2 N}{4J z} \biggl[
2 \left(\frac{\omega_u^2}{\overline{\omega}} + \overline{\omega} \right) \\
- \frac{\pi (\gamma H)^2 \omega_p^2}{2 \overline{\omega}^3}
\int_{-\infty}^{+\infty}
\frac{d k}{(1 + k^2 + \omega_p^2 / \overline{\omega}^2) 
\cosh^2(\pi k / 2)} \biggr] \;.
\end{multline}
If $\omega_p = 0$, the minimum of the action \eqref{ea-action0} is reached at
$\overline{\omega} = \omega_u$ and we return to Eq.~\eqref{rotator-action}
again.  Thus, in agreement with the previous analysis the Euclidean action
depends on the field for $\omega_p \neq 0$ only.  If the ratio $\omega_p /
\omega_u$ is small, this dependence is weak, and for the case of the field
directed along the easy axis the Euclidean action is
\begin{equation} \label{ea-action}
\mathcal{A}_E(EA) =
2 \hbar N S \sqrt{\frac{B_u}{J z}} \left[
1 - \frac{\pi (\gamma H)^2 \omega_p^2}{2 \omega_u^4} + \dots \right] \;.
\end{equation}
Thus, for $\omega_p \ll \omega_u$ the dependence of the Euclidean action
$\mathcal{A}_E$ on the field is weaker than it can be obtained from the naive
consideration (see Fig.~\ref{f:ea-action}).  The problem also can be solved in
the limit case $\omega_p \gg \omega_u$, when the approximate solution of
Eq.~\eqref{ea-operator} can be written as
\begin{equation}
l_y \simeq
\frac{2 i \gamma H \overline{\omega}}{\omega_p^2}
\frac{\sinh\xi}{\cosh^2\xi}  \ll 1  \;.
\end{equation}
The field dependence of the Euclidean action can be evaluated as $\mathcal{A}_E
\sim B_u(H) = \sqrt{1 - (H / H_u)^2}$.  Thus, the naive consideration of the
magnetic field through the renormalization of the anisotropy is recovered only
in the limit case of high planar anisotropy $\omega_p / \omega_u \to \infty$.

\begin{figure}
\includegraphics[bb = 65 510 430 775, width = \figwidth]{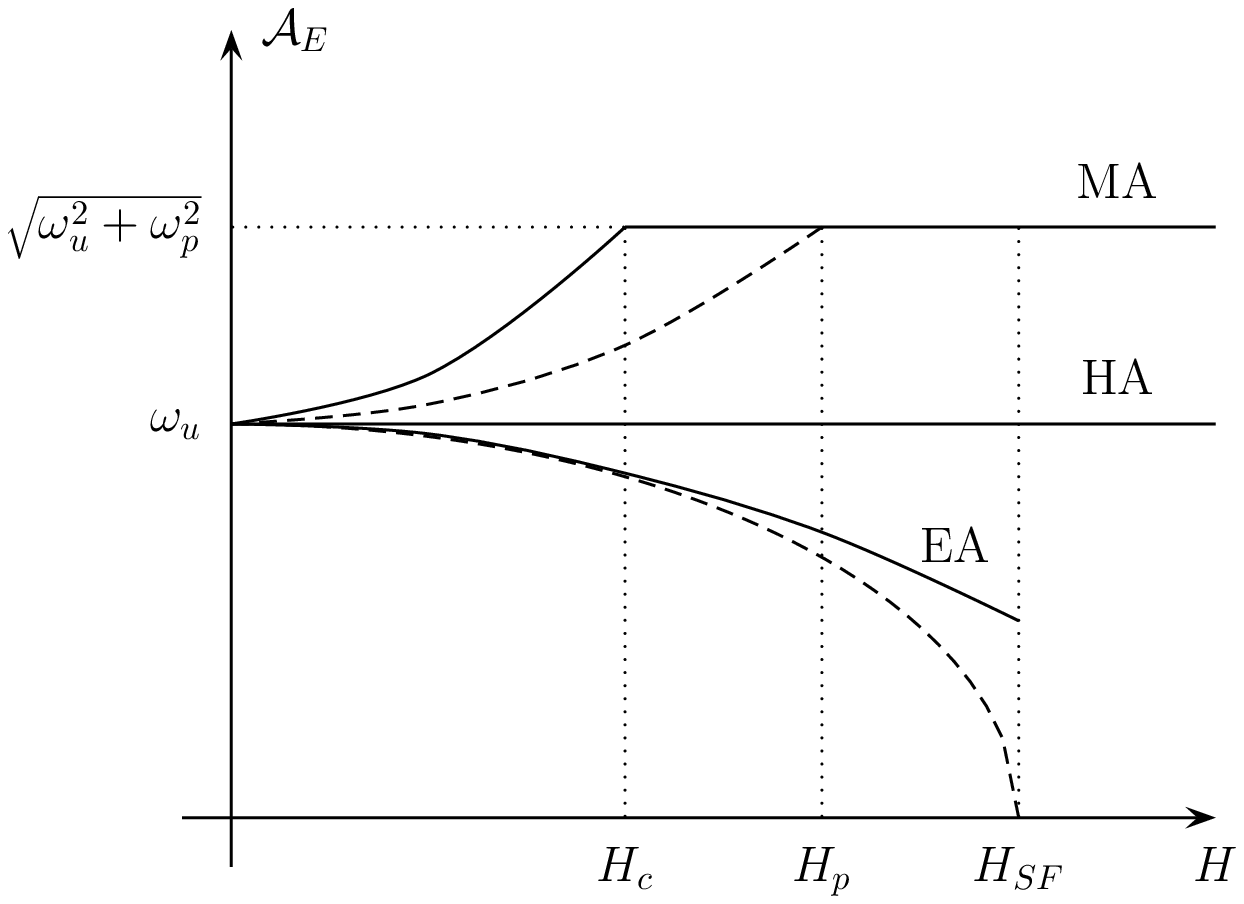}
\caption{The real part of the Euclidean action in units of $\hbar^2 N
  / 2J z$ as a function of the magnetic field (schematically).  Data for the
  field directed along the easy, medium, and hard axes are marked as EA, MA and
  HA, respectively, close to the appropriate curves.  Solid lines represent the
  calculations of $\mathcal{A}_E$ on the base of the full system~\eqref{eul},
  dashed lines are results of the naive consideration.}
\label{f:ea-actions}
\end{figure}

\subsection{Field perpendicular to the easy axis}

The case of the field directed along the hard axis is the simplest one.  The
plane $xz$ remains preferable for the rotation of the vector $\bm{l}$ for all
values of the field.  The right-hand sides of the system~\eqref{eul} are
proportional to $ \sin^2\theta \sin\phi$, and the exact solution of the
system~\eqref{eul} is $\phi = \phi_0 = \pi k$, $\overline{\omega} = \omega_u$.
It corresponds to the rotation in the most preferable plane $xz$.  The real part
of the Euclidean action is independent of the field, but the imaginary parts
have opposite signs for the two equivalent instanton trajectories with $\phi =
0$ and $\phi = \pi$,
\begin{equation} \label{ha-action}
\mathcal{A}_E(HA) =
2 \hbar N S \sqrt{\frac{B_u}{J z}} \pm
i\frac{\pi \hbar g \mu_B H N}{2 J z} \;.
\end{equation}
The nonzero imaginary part of the action~\eqref{ha-action} leads to interference
of the instanton trajectories and an oscillating dependence of the transition
probability on the field.\cite{GolPopkov95, ChiolLoss98} In this geometry the
most interesting effect arises due to the fluctuation
determinant.\cite{ChiolLoss98, Hu+00} The equations for small fluctuations
$\vartheta$ and $\mu$ around the instanton solution are uncoupled again, but a
complex-valued potential appears in the equation for $\vartheta$.  The
eigenvalues and the fluctuation determinant are also complex.  Thus, the
presence of $D_\theta$ changes not only the tunneling amplitude for the
instanton solution with a given value of $\sigma$, but also a phase shift for
two instanton paths with $\sigma = \pm 1$.  At low fields $D_\theta$ is almost
real.  It creates a decrease of the tunnel amplitude with the growth of the
field.  When the field increases, the factor $D_\theta$ produces the phase shift
of oscillations of $\Delta(H)$ caused by the interference.  The authors of
Ref.\cite{Hu+00} suggest that such a complicated behavior of $\Delta(H)$ caused
by the fluctuation determinant $D_\theta$ and the interference of instanton
trajectories with $\im\mathcal{A}_E \neq 0$ can be a universal feature for a
wide class of antiferromagnets.  As we saw earlier, for the field directed along
the easy axis the dependence $\Delta(H)$ is strongly different from the scheme
proposed by the authors of Ref.\cite{Hu+00}. For the field directed along the
medium axis the behavior is also essentially different.

The case of the field parallel to the medium axis is the most complicated one,
but it is interesting.  The right-hand sides of Eqs.~\eqref{eul} are
proportional to $\sin^2\theta \cos\phi$.  One type of exact solutions with the
rotation of $\bm{l}$ in the plane $xy$ containing the hard axis $y$ (a planar
solution) can be written as $\phi = \phi_0 = \pi (k + 1/2)$ and $\cos\theta =
\tanh (\overline{\omega}\tau)$ with $\overline{\omega} = \sqrt{\omega_u^2 +
  \omega_p^2}$.  For this planar solution the structure of the Euclidean action
$\mathcal{A}_E^{(p)}(MA)$ is the same as for the case of the field directed
along the hard axis.  Particularly, the real part of $\mathcal{A}_E^{(p)}(MA)$
does not depend on the field and the imaginary part is proportional to the
field
\begin{equation} \label{ma-action}
\mathcal{A}_E^{(p)}(MA) =
2 \hbar N S \sqrt{\frac{B_u + B_p}{J z}}
\pm i \frac{\pi \hbar g \mu_B H N}{2 J z} \;.
\end{equation}
For such instanton trajectories the fluctuation determinant is similar to the
factor obtained for the case of the field directed along the hard axis.  It
contains an imaginary part which leads to an extra contribution to the
interference effects.  But in contrast to the case of the field directed along
the hard axis here the plane $\phi = \pi (k + 1/2)$ is not the plane passing
through the medium axis.  The solution with $\phi = \phi_0 = \pi (k + 1/2)$ and
$\overline{\omega} = \sqrt{\omega_u^2 + \omega_p^2}$ satisfies the
system~\eqref{eul} exactly, but the real part of the action is not minimal, at
least at small magnetic fields $H < H_p$.  In order to explain this fact it is
sufficient to consider the value $H = 0$, when we have an exact solution with
the rotation in the plane $z x$ $\phi = \pi k$ and a smaller real part of the
action.  Thus, only for high fields the planar solution can be relevant.  For
the naive consideration of the field the real part of the Euclidean action for
the solution with $\phi = \pi k$ and $\theta = \theta(\tau)$ has to be
proportional to $\sqrt{\omega_u^2 + (\gamma H)^2}$, and
$\re\mathcal{A}_E^{(p)}(MA)$ is equal to $\re\mathcal{A}_E^{(p)}(HA)$ at the
point $H = H_p$.  But as we will see below, the situation is actually more
complicated.

Due to nonzero gyroscopic terms in the system~\eqref{eul} an exact solution with
$\phi = \pi k$ does not exist for the case $H_x \neq 0$ and an appropriate
approximate instanton solution is nonplanar.  Following the treatment of the
preceeding section, we write the solution as $\cos\theta = \tanh
(\overline{\omega} \tau)$, $l_y = \sin\phi \sin\theta \ll 1$ and determine $l_y$
from the linear equation
\begin{equation} \label{ma-operator}
\left(\widehat{M} + \epsilon \right) l_y(\xi) =
\frac{2i \gamma H}{\overline{\omega}} \;
\frac{1}{\cosh^2\xi} \;,
\end{equation}
where $\epsilon = [\omega_p^2 - (\gamma H)^2] / \overline{\omega}^2$, $\xi =
\overline{\omega} \tau$, and $\widehat{M}$ is defined by Eq.~\eqref{BOSH}.  In
contrast to the similar equation~\eqref{ea-operator} the right-hand side of
Eq.~\eqref{ma-operator} is symmetric with respect to $\xi$ and a contribution
from the localized eigenfunction~\eqref{BOSH-spectrum-loc} is present.  This
contribution is proportional to $1 / \epsilon$ and is mostly important for the
case $\gamma H$, $\omega_p \ll \omega_u$.  As we will see below, such an
instanton is important at low field and the abovementioned restriction is
irrelevant.  The Euclidean action for this nonplanar instanton as a function of
$\overline{\omega}$ at $\omega_p < \omega_u$ can be written as
\begin{multline} \label{ma-action0}
\mathcal{A}_E(\overline{\omega}) =
\frac{\hbar^2 N}{4J z} \biggl[
2 \frac{\omega_u^2 + \gamma^2 H^2}{\overline{\omega}} + 2 \overline{\omega} \\
+ \frac{(\pi \gamma H)^2 \overline{\omega}}{2 (\omega_p^2 - \gamma^2 H^2)}\biggr] +
\Delta \mathcal{A}_E \;,
\end{multline}
where $\Delta \mathcal{A}_E$ is a contribution from the continuous spectrum that
is determined through an integral over $k$ with a structure that similar to
Eq.~\eqref{ea-action0}.  This term contains the factor $(\gamma H /
\overline{\omega})^2$ and it can be omitted for the case of interest $\gamma H
\sim \omega_p \ll \overline{\omega}$.  After minimizing over $\overline{\omega}$
the action for a nonplanar instanton for the case $\omega_p < \omega_u$ takes
the form
\begin{equation} \label{ma_h-action}
\mathcal{A}_E^{(np)}(MA) =
2 \hbar N S \sqrt{\frac{B_u (1 + H^2 / H_u^2)}{J z}}
\sqrt{1 + \frac{\eta H^2}{H_p^2 - H^2}} \;,
\end{equation}
where numerical constant $\eta = \pi^2 / 4$.  The problem can be solved in the
opposite limit case, namely, at $\omega_u < \omega_p$.  For this case one can
replace the operator $\widehat{M} + \epsilon$ by $\epsilon$, and the approximate
solution reads $l_y(\xi) = 2i \gamma H / (\epsilon \overline{\omega}
\cosh^2\xi)$.  Then, for the Euclidean action we arrive to the same
equation~\eqref{ma_h-action}, but with another numerical constant $\eta = 8/3
\approx 2.667$.  Comparing this value found for $\omega_u < \omega_p$ with
$\pi^2 / 4 \approx 2.467$ for $\omega_p < \omega_u$, we can tell that for the
two opposite limit cases the Euclidean action $\mathcal{A}_E^{(np)}(MA)$ is
approximately described by the same equation.  Thus, we can suggest the
Eq.~\eqref{ma_h-action} is a good approximation for any relation between
$\omega_p$ and $\omega_u$, that is in line with numerical data, see
Sec.~\ref{s:numerics}.

Thus, when the field is parallel to the medium axis, the Euclidean action is
real for a nonplanar instanton, and due to Eq.~\eqref{ma_h-action} it increases
faster than it can be expected from the static renormalization of the anisotropy
constant.  It is possible to show that both fluctuation determinants $D_\theta$
and $D_\phi$ are also real.  The interference effects are absent, and the tunnel
splitting monotonically decreases with the growth of the field at small $H <
H_c$, where $H_c$ is a critical field, at which the values of the Euclidean
action for the nonplanar instanton becomes equal to the real part of
$\mathcal{A}_E^{(p)}(MA)$ for planar instantons.  For small anisotropy in the
basal plane, $\omega_p \ll \omega_u$, the value of $H_c$ is small, and
\begin{equation}\label{redir-field}
H_c =
H_p \frac{B_p}{B_p + (\pi /2)^2 B_u} \ll H_p \;.
\end{equation}
But it is smaller that $H_p$ even in the opposite limit case $\omega_u \ll
\omega_p$,
\begin{equation}\label{redir-field-num}
H_c =
H_p \sqrt{\frac{1}{1 + \sqrt{\eta}}} \approx 0.6163 H_p \;.
\end{equation}
Thus, at low magnetic fields $H < H_c$ the Euclidean action for the nonplanar
instanton solution $\mathcal{A}_E^{(np)}(MA)$ is lower than for the planar
solution.  Its value reaches $\mathcal{A}^{(p)}_E(MA)$ at $H = H_c$, and the
scenario of tunneling is changed to the planar one, common to that is present
for the field along the hard axis with the tunneling exponent independent of the
magnetic field and with interference effects caused by imaginary parts of both
$\mathcal{A}_E$ and fluctuation determinant.


\section{Numerical data \label{s:numerics}}

The semiclassical analysis of the coherent quantum tunneling between the
classically degenerated ground states demonstrates that the level splitting is
highly sensitive to the orientation of the magnetic field.  Among the considered
field orientations along the axes of rhombic symmetry the cases of the easy and
medium axis are the most interesting ones.  In both cases the Euclidean action
has a zero imaginary part, and the corresponding interference effects are
absent.  We also have shown that for these two cases the preexponential factor,
which could be a source of interference, is real.  Thus, for these orientations
of the magnetic field interference effects does not appear, and the level
splitting is mainly determined by the dependence of the real part of the
Euclidean action on the magnetic field $H$.  The character of this function is
determined by the anisotropy $B_p$ in the basal plane.  The dependence is absent
for the case $B_p = 0$ and the field parallel to the easy axis only.  The
exponential factor $\exp(-\mathcal{A}_E / \hbar)$ is an increasing function of
$H$ for the field parallel to easy axis, and a decreasing function of $H$ for
the field parallel to the medium axis.  This behavior strongly differs from that
is present for the field directed parallel to the hard axis.

In order to check the semiclassical results found in assumption of some
inequalities such as $B_p \ll B_u$ or $B_u \ll B_p$ and to estimate the role of
the fluctuation determinant, which was not investigated here, we diagonalize
numerically\footnote{The implicitly restarted Arnoldi method implemented in the
  ARPACK package of numerical routines is used for eigenvalue calculations of
  large scale sparse matrices.} the Hamiltonian~\eqref{Hamiltonian} for the
two-spin quantum model with high enough values of the spin (up to $S = 100$), a
small uniaxial anisotropy $B_u / J = 0.01$ -- $0.1$ that guarantees reasonable
values of the level splitting, and for several values of $B_p / B_u$, see
figures below in this section.  The numerical results show a satisfactory
agreement with the instanton approximation even for moderate values of the spin
$S = 10$ -- $20$ even without taking into account the preexponential factor.
But some discrepancies, which cannot be attributed to the approximations used in
the analytical consideration, are also seen.

\begin{figure}
\includegraphics[bb = 70 450 530 800, width = \figwidth]{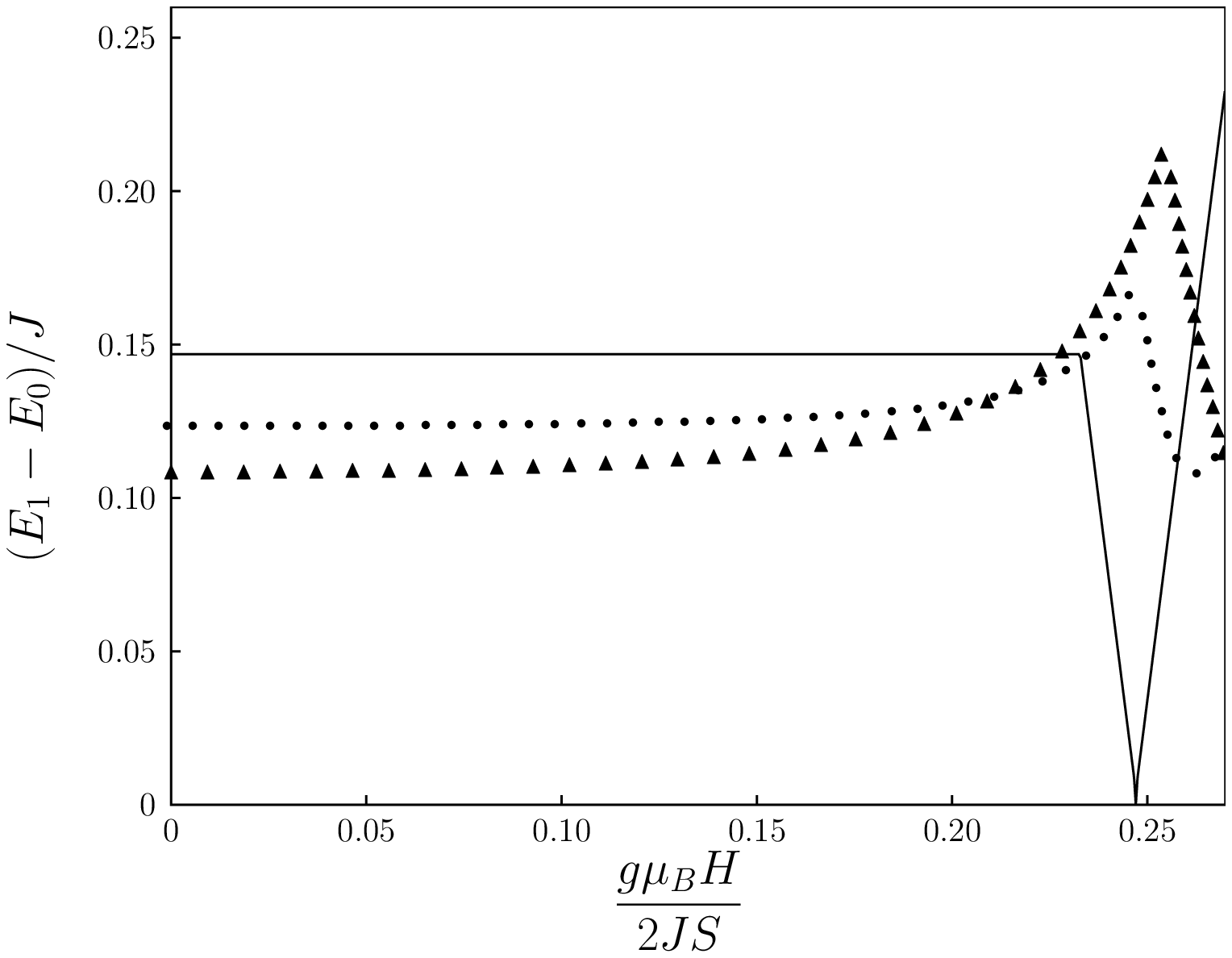}
\caption{Splitting of the lowest level for the quantum model with
  Hamiltonian~\eqref{Hamiltonian}, spin $S = 5$ and $B_u / J = 0.1$ for three
  values of the ratio $B_p / B_u = $ 0.0 (solid lines), 0.2 (circles), 0.4
  (triangles).  The magnetic field is directed along the easy axis and
  normalized to the exchange field $H_{ex} = J z S N / (g \mu_B)$ with $z = 1$
  and $N = 2$.  The cusps on the curves correspond to the change of the ground
  state that is a quantum counterpart of the spin-flop transition.}
\label{f:3x3-e}
\end{figure}

For the case of the field directed along the easy axis at $B_p = 0$ the tunnel
splitting is independent of the field up to the point of the spin-flop
transition.  This behavior is exactly reproduced by means of a numerical
diagonalization, see Fig.~\ref{f:3x3-e}.  Due to the results of the instanton
approach, for $B_p \neq 0$ the Euclidean action decreases, and the level
splitting increases with the growth of the field $H$.  The effect becomes more
pronounced at large $B_p$.  This behavior well corresponds to the numerical data
represented in Fig.~\ref{f:3x3-e}.  The only difference between the numerical
data for the level splitting $E_1 - E_0$ and the semiclassical results is that
the field of the spin-flop transition identified as a field of the cusp on
curves in Fig.~\ref{f:3x3-e} slowly depends on the ratio $B_p / B_u$ for the
quantum model.  This dependence is completely absent in the semiclassical
approximation, in particular, in the instanton approach.  Probably, close the
point of the spin-flop transition, where the sharp decrease of the Euclidean
action is present, the value of $\mathcal{A}_E / \hbar$ becomes comparable with
unity even for large $S$, and the quantum fluctuations treated beyond the
semiclassical approximation become important.

\begin{figure}
\includegraphics[bb = 70 355 550 800, width = \figwidth]{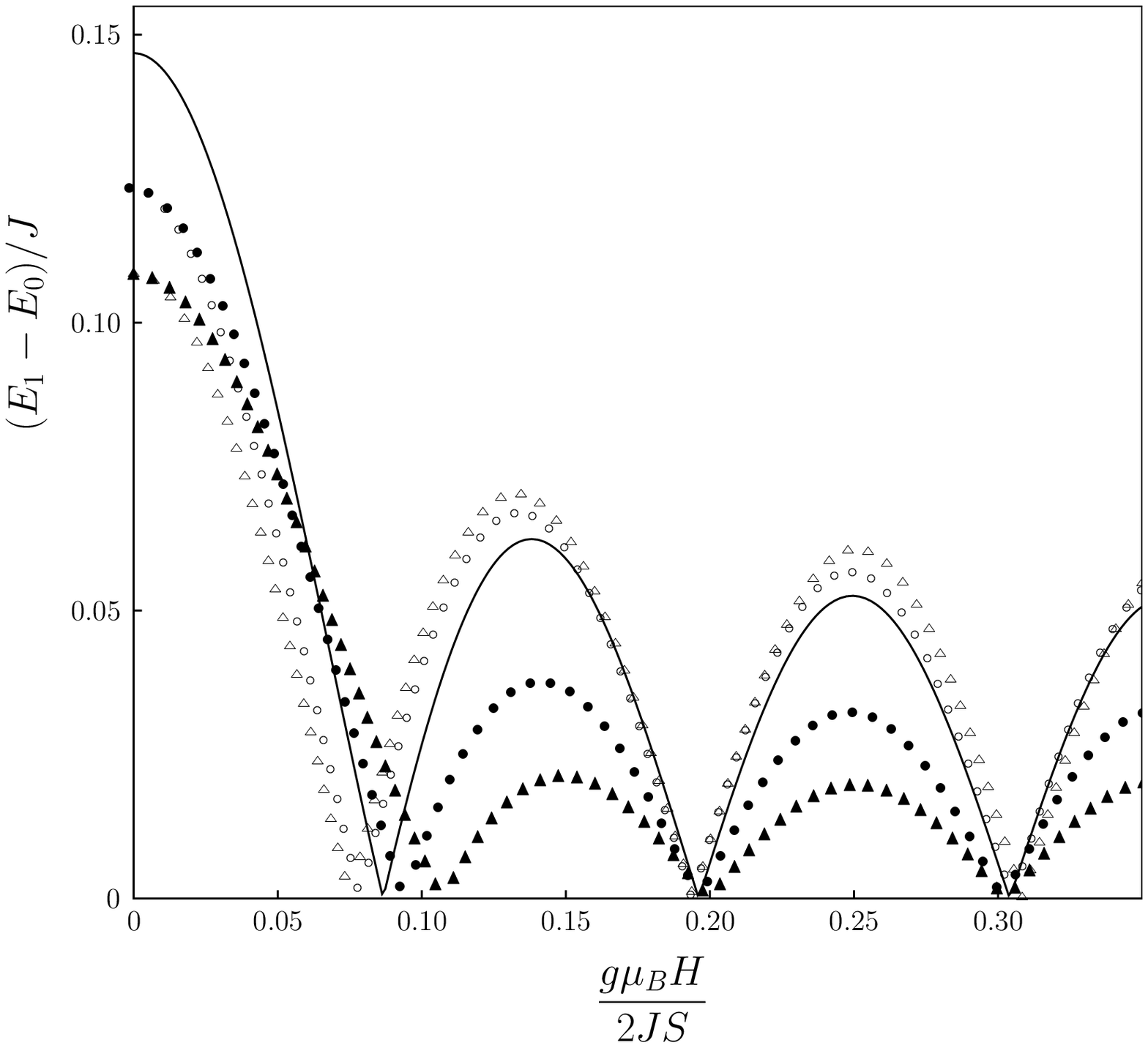}
\caption{Splitting of the lowest level for the quantum model with
  Hamiltonian~\eqref{Hamiltonian}, spin $S = 5$, and $B_u / J = 0.1$ for three
  values of the ratio $B_p / B_u = $ 0.0 (solid lines), 0.2 (circles), 0.4
  (triangles), and the magnetic field directed along the hard axis (open
  symbols) or medium axis (full symbols).  The field is normalized to the
  exchange value.}\label{f:3x3-mh}
\end{figure}

In the case of the field directed along the hard axis as well as in the case of
high fields, $H > H_c$, along the medium axis the instanton approach predicts
that the tunneling occurs through the planar instanton paths.  For these cases
the level splitting $\Delta(H)$ oscillates as a function of the field with a
constant period.  If the field is directed along the hard axis, the behavior of
$\Delta(H)$ coincides with the results of Refs.\cite{ChiolLoss98, Hu+00}.  The
amplitude of oscillations is determined by the preexponential factor only, and
it weakly depends on $B_p$.  This kind of behavior is clearly seen in
Fig.~\ref{f:3x3-mh}.  In the case of high fields, $H > H_c$, directed along the
medium axis the tunneling is also determined by planar instantons with a nonzero
imaginary part of the action and interference effects appear.  But when the
ratio $B_p / B_u$ increases, the main exponential factor drastically decreases,
and the amplitude of oscillations $\Delta(H)$ decreases also.  This feature is
in a good agreement with the numerical calculations depicted in
Fig.~\ref{f:3x3-mh}.

At low fields, $H < H_c$, parallel to the medium axis the tunneling is
determined by nonplanar instantons.  The real part of the Euclidean action $\re
\mathcal{A}^{(np)}_E(MA)$ monotonically increases, and the instanton approach
predicts a strong monotonic decrease of $\Delta(H)$ up to $H = H_c$.  For any
values of $B_p / B_u$ at $H < H_c$ the instanton approach also predicts that
oscillations caused by interference are absent.  The effect of the axes
reorientation in the basal plane at $H = H_p$ does not clearly manifest itself
as the spin-flop transition in Fig.~\ref{f:3x3-e}, but it can be understood as a
shift of a point, where oscillations start, to the region of high fields.  Both
factors, as well as the growth of the characteristic field of transition to the
high-field tunneling picture $H_c$, are in a qualitative agreement with data of
numerical calculations presented in Fig.~\ref{f:3x3-mh}.

\begin{figure}
\includegraphics[bb = 65 500 485 780, width = \figwidth]{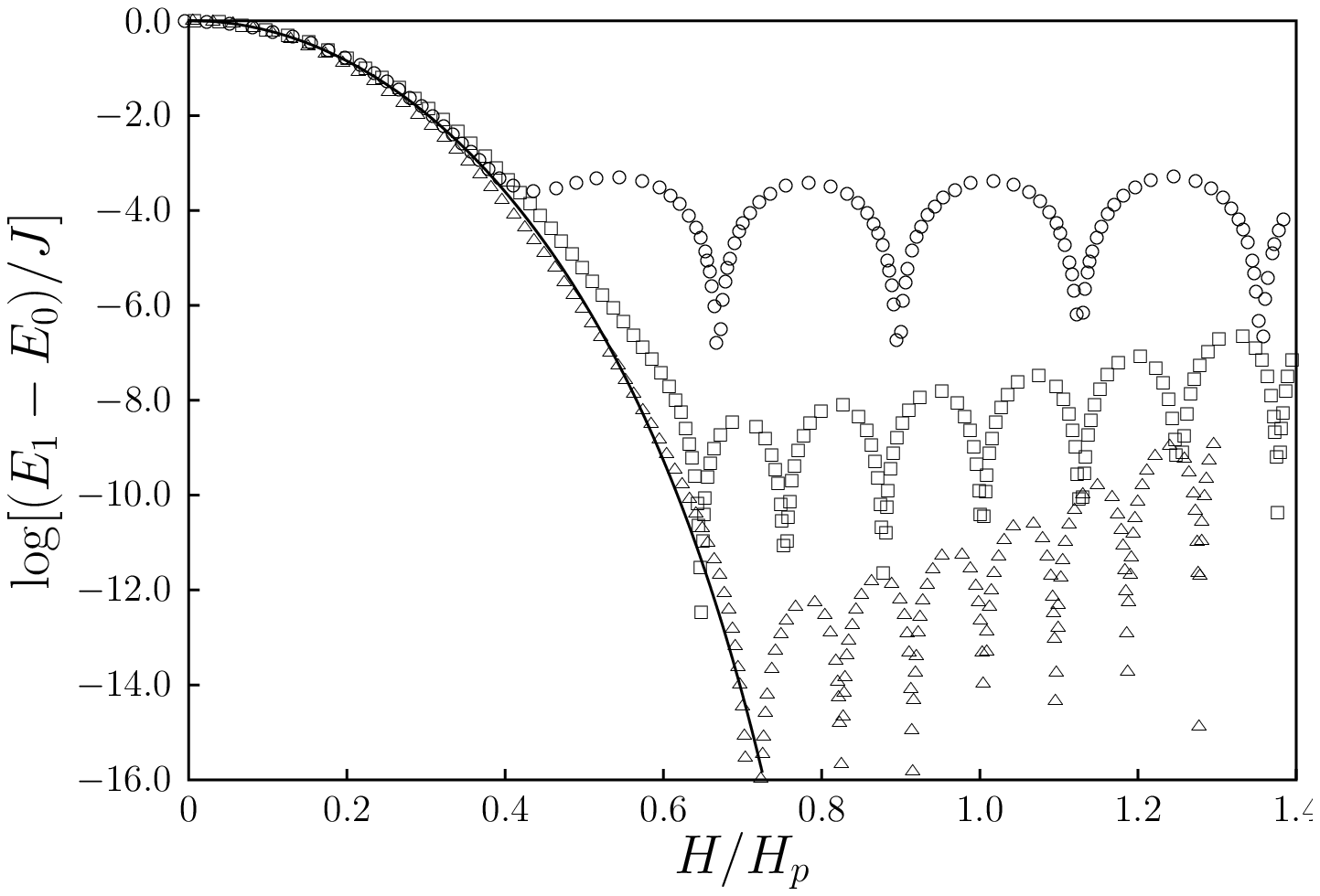}
\caption{Numerical data for the logarithm of the level splitting
  normalized by the value at $H = 0$ versus the magnetic field directed along
  the medium axis for the spin $S = 10$ and $B_u / J = 0.1$ and three values of
  the anisotropy constant in the basal plane: $B_p / J = $ 0.05 (circles), 0.15
  (squares), 0.25 (triangles).  The magnetic field is normalized to the
  reorientation field $H_p$.  The solid line describes the theoretical
  dependence~\eqref{ma_h-action} for a nonplanar instanton.}
\label{f:3x3-m}
\end{figure}

In order to give a more detail comparison of the analytical and numerical data
for small fields, $H < H_c$, parallel to the medium axis, we investigated
numerically the value of spin $S = 10$, for which the role of the field
dependence of the fluctuation determinant is expected to be less important.  The
data together with the simple theoretical estimate of the level splitting, see
Eq.~\eqref{ma_h-action}, in the form $\Delta(H) / \Delta(0) =
\exp\{-[\mathcal{A}(H) - \mathcal{A}(0)] / \hbar\}$, where $\mathcal{A}(H)$ is
the value of the Euclidean action for nonplanar instantons, are present in
Fig.~\ref{f:3x3-m}.  Here we can say about at least a semiquantitative
agreement.

\begin{figure}
\includegraphics[bb = 110 495 535 775, width = \figwidth]{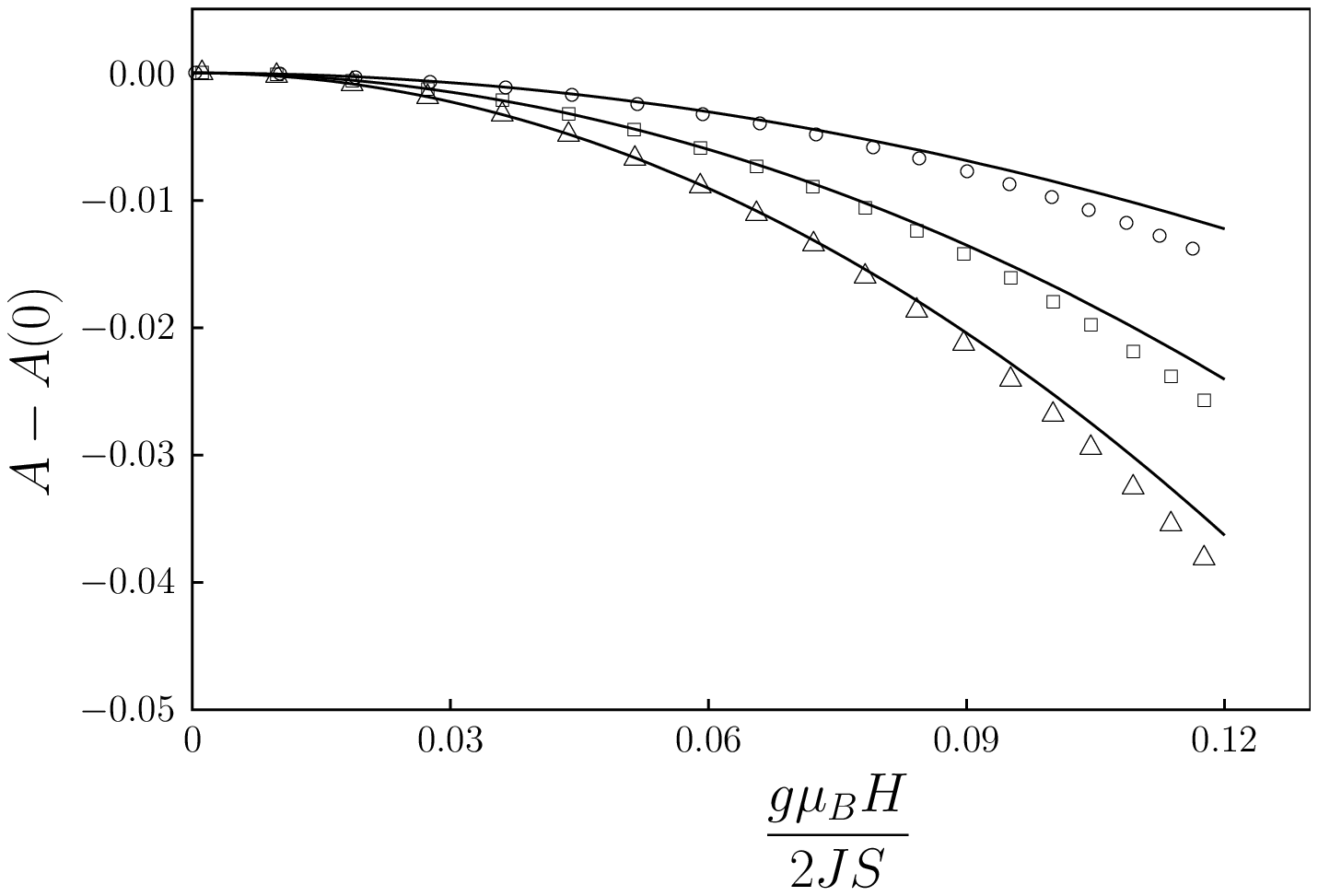}
\caption{Real part of the Euclidean action per unit spin obtained
  from the quantum model with Hamiltonian~\eqref{Hamiltonian}, $B_u / J = 0.1$
  for three values of the ratio $B_p / J = $ 0.02 (circles), 0.04 (squares),
  0.08 (triangles) and the magnetic field directed along the easy axis.  Solid
  lines are analytical predictions, Eq.~\eqref{ea-action}.  The field is
  normalized to the exchange value.}
\label{f:ea-action}
\end{figure}

\begin{figure}
\includegraphics[bb = 120 500 520 790, width = \figwidth]{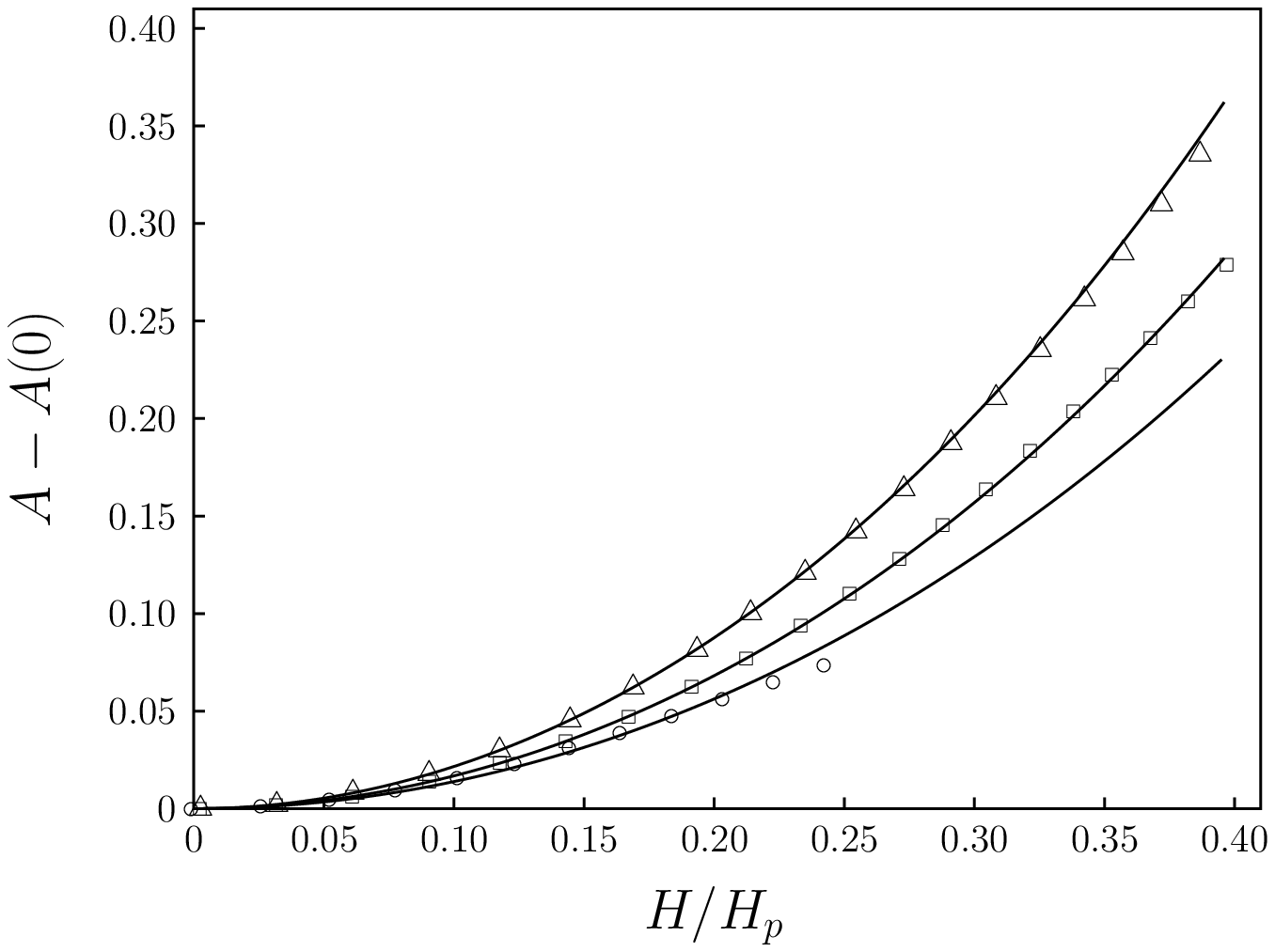}
\caption{Real part of the Euclidean action per unit spin obtained
  from the quantum model with Hamiltonian~\eqref{Hamiltonian}, $B_u / J = 0.1$
  for three values of the ratio $B_p / J = $ 0.05 (circles), 0.1 (squares), 0.2
  (triangles) and the magnetic field directed along the medium axis.  The
  magnetic field is normalized to $H_p$ for each value of $B_p$.  Solid lines are
  analytical predictions, Eq.~\eqref{ma_h-action} with $H_p$ as a trial
  parameter.}
\label{f:ma-action}
\end{figure}

In order to check quantitatively the analytical expressions~\eqref{ea-action}
and \eqref{ma_h-action} for the real part of the Euclidean action, we propose
the method of extracting the value of $\mathcal{A}_E$ from the level splitting
data without using an explicit expression for the preexponential factor.  This
method is based on the general theoretical formula for the level splitting
\begin{equation}\label{fit3}
\Delta = D S^\beta \exp(- SA) \;,
\end{equation}
where $D$, $\beta$ and $A$ are functions independent of the spin with normalized
values of the magnetic field $H / H_p$ or $H / H_u$.  In the semiclassical
approximation the function $A$ means the real part of the action per unit of
spin, $D$ is the preexponential factor and $\beta$ is associated with the number
of zero modes: $\beta = 2$ for the case of the purely uniaxial symmetry, that is
$B_p = 0$ and the magnetic field parallel to the easy axis, see also
Eq.~\eqref{Delta} and discussion there; $\beta = 3/2$ for all rest models with
biaxial symmetry.

The parameter $\beta$ is also chosen floating in order to absorb corrections to
the $\sigma$-model and its semiclassical treatment.  This method could, in
principle, be applied for any semiclassical problem, but it is mostly useful for
problems, where the imaginary part of the action, as well as the preexponential
factor, are zero, and interference effects with oscillations are absent.  For
the tunneling in the external field this is just the problem we are interested
in.  This condition is realized at $H < H_u$ for the field directed along the
easy axis and $H < H_p$ for the case of the field parallel to the medium one.

In order to obtain the values of interest, a numerical calculation of splitting
of the lowest level is performed for few (at least, three) large values of spin.
The real part of the action can be easily extracted from the numerical data such
as qualitatively presented in Figs.~\ref{f:ea-action} and \ref{f:ma-action}.
The functions $A_E(H)$ in these figures are obtained by fitting the numerical
data for the spin $10 < S < 20$ and each value of the field using
Eq.~\eqref{fit3}.

For the simplest uniaxial case $B_p = 0$ and the field directed along the easy
axis we found that the expected value $\beta = 2$ appears, and the Euclidean
action is independent of the field up to the point of the spin-flop transition.
Then, with growth of $B_p$ we found a very sharp transition to the regime with
one zero mode, for which the value of $\beta = 3/2$ is reproduced.  Thus,
checking that $\beta$ does not differs significantly from $3/2$ (the range $1.5
\pm 0.1$ is taken) side by side with more trivial conditions that the obtained
values of $D$, $\beta$ and $A$ are stable with respect to varying $S$ and the
exponent $S A$ is large enough, we select sets of the parameters for the quantum
model~\eqref{Hamiltonian}.

The action $A$ obtained by fitting of Eq.~\eqref{fit3} with the fixed value
$\beta = 1.5$ is plotted as functions $A(H)$ in Figs.~\ref{f:ea-action} and
\ref{f:ma-action} for the fields directed along the easy and medium axes.  Note
that the action $A(0)$ at zero field is subtracted from the functions $A(H)$.
The analytical theory gives the value $A(0) = 4\sqrt{B_u / J}$ independent on
$B_p$, but numerical calculations demonstrate a weak dependence of $A(0)$ on
$B_p$.  In details, $A_{th}(0) = 1.26$ for $B_u / J = 0.1$ and $A_{num}(0)$ is
1.22, 1.19 and 1.14 for the same value $B_u / J = 0.1$ and $B_p / J$ equal to
0.05, 0.10, 0.20, respectively.

The only way to describe the numerical data is to consider the fields $H_u$ and
$H_p$ in Eqs.~\eqref{ea-action} and \eqref{ma_h-action} as phenomenological
parameters that predicted by the classical expressions~\eqref{SF-field} and
\eqref{redir-field}.  For simplicity, we rescale the fields in
Fig.~\ref{f:ma-action} as $H \to H / H_p$.  The numerical data are fitted by
Eqs.~\eqref{ea-action} and \eqref{ma_h-action} using these fields in the range
$H < 0.1 H_u$ and $H < 0.2 H_p$.  Obtained values for a trial parameter
$\tilde{H_p} / H_p = $ 1.164, 1.144, 1.145 for $B_p / J = $ 0.05, 0.10, 0.20 are
in a good agreement with the classical result, where obviously $\tilde{H_p} =
H_p$.  Appropriate analytical curves are plotted as solid lines in
Figs.~\ref{f:ea-action} and \ref{f:ma-action}.  In the case of the medium axis
the theory predicts the shape of the curves obtained from the quantum model with
a good accuracy up to fields close to $H_c$.  For the easy axis the action
decreases more significantly that it would be expected from the perturbative
treatment of the semiclassical model.  In both cases we can pretend on the
quantitative agreement of the proposed theory and numerical data.

It is important to note one more discrepancy between the developed analytical
theory and the presented numerical data.  The analytical expression for the
level splitting in the field directed along the hard axis does not contain any
dependency on $B_p$, but in Fig.~\ref{f:3x3-m}, as well as in the numerical data
of Refs.\cite{ChiolLoss98, Hu+00}, this dependency is present.  It is more
important for higher values of $B_p / B_u$.  To explain it, as well as the
observed dependence of the parameters $H_p$ and $H_u$, we note that the
$\sigma$-model treats antiferromagnets in the first approximation over small
ratios of the anisotropy constants or the magnetic field to the exchange
integral $J$.  Here, the values of these ratios was taken in the range 0.1 --
0.2, and the deviation of the $\sigma$-model results that is of order of 10\%
from the numerical calculations are not surprising.


\section{Concluding remarks \label{s:conclusion}}

In conclusion, the antiferromagnetic particles can show a reach variety of
tunneling behaviors that depend on the direction of the magnetic field.  In
addition to the oscillation behavior for the field directed along the hard
axis,\cite{ChiolLoss98, Hu+00} we found the growth of the tunnel splitting
$\Delta$ for the field directed parallel to the easy axis and a steep decrease
of $\Delta$ for the field along the medium axis.  Both mentioned behaviors are
connected to the tunnel exponent dependence on the field.  It is important to
note that such effects cannot be directly associated with the decrease or
increase of the tunnel barrier, respectively, that governed by the static
renormalization of the anisotropy energy.

Let us briefly discuss the possibility for experimental investigations of the
tunneling effects predicted in the paper.  The main point consists in kinds of
antiferromagnets that could be used for experiments.  The traditional
antiferromagnetic samples such as small ferritin particles have unpaired spins
and behave as to noncompensated antiferromagnets.  For this reason the
destructive interference for them is mainly dictated by the excess spin in the
way common to ferromagnets.\cite{Chudn95, ChiolLoss97} Moreover, the presence of
a nonzero total magnetic moment drastically changes the structure of the ground
state.  It is enough to say that the degeneracy is absent except some fixed
directions of the field with respect to the crystalline axis.\cite{GolPopkov95,
  GolPopkov95jetp} We proposed a way to overcome this problem,\cite{IvKir99} but
limitations caused by noncompensated spins seams to be more serious.  Note that
the same problem appears for ferromagnetic particles where the effects of the
barrier reduction\cite{ChudnGunter88prl} and the oscillation behavior of the
ground-state tunnel splitting\cite{Garg93, Garg99prb} was predicted many years
ago, but observed only recently.\cite{WernsSess99}

The key point in this important experimental success is based on the synthesis
of high-spin molecules packed in the well-oriented monocrystals.  Up to our
understanding, the first possibility to investigate purely antiferromagnetic
features is to use high-spin molecules with a well-defined spin structure.  The
molecules with ferromagnetic and antiferromagnetic couplings, uniaxial and
rhombic anisotropies have been synthesized in the recent
years.\cite{Wernsdorfer01} For known ferromagnetic molecules such as Fe$_8$ the
splitting is small, but the technique developed by Wernsdorfer and
Sessolli\cite{WernsSess99} allows one to measure a very small tunnel splitting
of order of $10^{-8}$K.  The first possibility discussed by many authors
consists in using spin rings with antiferromagnetic coupling.  For well-known
antiferromagnetic molecular magnets such as Fe$_{10}$, Fe$_6$, V$_8$ the problem
is opposite to that for ferromagnetic molecules: the anisotropy is too small,
and the barrier is too low to see clear semiclassical effects such as
MQT.\cite{MeierLoss01prb, Normand+01} On the other hand, antiferromagnetic rings
of eight chromium ions with a high anisotropy have been recently
synthesized.\cite{vanSlageren+02} One more possibility is to use spin dimers
containing two coupled high-spin molecules (molecular magnets) with a
ferromagnetic coupling inside the molecule and an antiferromagnetic
intermolecular coupling.  For instance, the observation of the well-structured
dimers of high-spin molecules Mn$_4$ (spin $S = 9/2$) with the antiferromagnetic
coupling between two Mn$_4$ molecules has been recently
reported.\cite{Edwards+03} Quantum tunneling in the Mn$_4$ dimers was
investigated experimentally.\cite{Wernsdorfer+02nat} Such dimers of high-spin
molecules such as Fe$_8$ with the macroscopic spin $S = 10$ and well pronounced
rhombic anisotropy could be a good candidates for observation of the effects
considered in our paper.\footnote{Surely, the exchange interaction between these
  two molecules is antiferromagnetic, as it takes place in the case of Mn$_4$;
  we were not able to find the corresponding data in the literature.}  It is
worth to note also that the predicted possibility of enlarging the value of the
Euclidean action (to suppress the tunneling) by means of the magnetic field
directed parallel to the medium axis can be useful for following investigations.

\begin{acknowledgments}

  The authors thank A.~K.~Kolezhuk for fruitful discussions and help, and
  H.-J.~Mikeska for useful discussions.  This work was supported by
  Volkswagen-Stiftung, grant No.~I/75895.  The authors also thank Institute of
  Theoretical Physics (University of Hanover) for kind hospitality.

\end{acknowledgments}

\bibliography{paper}

\end{document}